# Amorphous complexions enable a new region of high temperature stability in nanocrystalline Ni-W

Jennifer D. Schuler [a], Olivia K. Donaldson [b], Timothy J. Rupert [a,b,*]
[a] Department of Chemical Engineering and Materials Science, University of California, Irvine, CA 92697, USA
[b] Department of Mechanical and Aerospace Engineering, University of California, Irvine, CA 92697, USA
*E-mail: trupert@uci.edu

**Abstract:**

Solute segregation is used to limit grain growth in nanocrystalline metals, but this stabilization often breaks down at high temperatures. Amorphous intergranular films can form in certain alloys at sufficiently high temperatures, providing a possible alternative route to lower grain boundary energy and therefore limit grain growth. In this study, nanocrystalline Ni-W that is annealed at temperatures of 1000 °C and above, then rapidly quenched, is found to contain amorphous intergranular films. These complexions lead to a new, unexpected region of nanocrystalline stability at elevated temperatures.

Nanocrystalline metals, commonly defined as having an average grain size less than 100 nm, have demonstrated a variety of beneficial characteristics including increased strength [1, 2], wear resistance [3-5], and fatigue lifetime [6]. These enhanced properties are largely attributed to the high volume of grain boundaries present in these materials [7] and make nanocrystalline metals promising candidates for use in extreme environments, but the same grain boundaries that generate these desirable characteristics in turn serve to frustrate their application. Nanocrystalline metals are highly susceptible to grain growth, driven by a desire to lower the energy penalty associated with the high volume fraction of these defects [8-15]. Grain growth can be simply pictured by looking at the velocity of a grain boundary (*v*) [16]:

$$v = M_{gb} \cdot \gamma_{gb} \cdot \kappa \qquad (1)$$

where $M_{gb}$ is the grain boundary mobility, $\gamma_{gb}$ is the grain boundary energy, and $\kappa$ is the local mean curvature of the boundary. From Equation 1, grain boundary velocity and therefore grain growth is a function of kinetic and thermodynamic components. A tremendous amount of recent research has shown that these factors can be influenced by alloying and selective doping, ultimately suppressing grain growth in a range of nanocrystalline metals [17-33]. Thermodynamically-driven stabilization is characterized by solute segregation to the grain boundary because it is energetically favorable for dopants to reside at interfaces compared to the crystalline interior. Kinetic solute drag on grain boundary migration can also occur when there is solute segregation [27, 28]. Such behaviors have been captured in various computational models [27, 28, 34, 35], which predict alloys that can sustain stabilized nanocrystalline structures, and have been confirmed experimentally in binary alloys such as Ni-W [36], W-Ti [35], and Fe-Zr [37]. Nanocrystalline stability generally has limitations and eventually breaks down with increasing temperature [36], where rapid coarsening is observed.

Grain boundary segregation can also induce thermodynamically-driven interfacial structural transitions that are dependent on the local grain boundary composition and system temperature [38, 39]. These distinct grain boundary structural states can be called complexions, with features that range from ordered atomic layers to wetting films being observed in the literature [40]. Amorphous intergranular films (AIFs) are a type of complexion consisting of dopant-enriched grain boundaries with nanometer-scale thicknesses and disordered structure [41]. AIFs form when the energy needed to create an amorphous region of a given thickness with two new crystalline-amorphous interfaces is less than the energy of the original grain boundary [42]:

$$\Delta G_{amorph} \cdot h + 2\gamma_{cl} < \gamma_{gb} \qquad (2)$$

where $\Delta G_{amorph}$ refers to the volumetric free energy penalty for an undercooled amorphous film at a given binary alloy composition, $h$ is the film thickness, $\gamma_{cl}$ is the excess free energy of the new crystalline-amorphous interfaces, and $\gamma_{gb}$ is the excess free energy of the original crystalline grain boundary. In effect, AIF formation is a form of grain boundary premelting, which is prohibitively difficult to observe in pure metals [43] but accessible for highly doped boundaries. In binary metallic alloys, Schuler and Rupert found that AIF formation is favorable when there is dopant segregation to the grain boundary, which can be predicted by a positive enthalpy of segregation ($\Delta H^{seg}$), coupled with a negative enthalpy of mixing ($\Delta H^{mix}$) [44]. One would also need to take the alloy to a temperature that is high enough for the boundary premelting transition to occur, which Luo and coworkers predicted to be at least 60-85% of the alloy melting temperature [45, 46]. AIFs fundamentally form to lower the grain boundary energy and have been observed to enhance nanocrystalline thermal stability, operating under the same principles as thermodynamically-driven stabilization to limit $v$ from Equation 1. For example, mechanically-alloyed nanocrystalline

Cu-Zr with AIFs remained nanostructured even after a week at 98% of the solidus temperature [47].

Based on the materials selection rules introduced above, Ni-W would be expected to form AIFs at sufficiently high temperatures, with $\Delta H^{seg} = 10$ kJ/mol and $\Delta H^{mix} = -3$ kJ/mol [30]. This alloy system has been studied extensively in the past, with the unique feature that the grain size can be tailored during electrodeposition by controlling the reverse pulse current and, therefore, the W content [48, 49]. Prior work has shown that nanocrystalline Ni-W remains stable up to ~500 °C, primarily due to W grain boundary segregation [36]. However, at higher temperatures, Ni-W alloys have been observed to undergo rapid grain growth. Because of this rapid grain growth, most studies of thermal stability only extend to temperatures up to ~900 °C (see, e.g., [36]). While it appears that solute segregation-enabled stabilization breaks down by these temperatures, 900 °C is only ~60% of the melting temperature, meaning it is roughly at the lower limit of potential AIF formation. While AIFs have not been observed in Ni-W to date, we suggest that most studies have not gone to high enough temperatures. Moreover, very rapid quenching would be needed to freeze these features into the microstructure. Thus, we hypothesize that AIFs will form in Ni-W at sufficiently high temperatures, with the secondary hypothesis that these complexions will impact grain growth at these elevated temperatures. In this paper, the grain boundary composition and structure, as well as the grain growth behavior of nanocrystalline Ni-W, was analyzed at temperatures as high as 1200 °C. We find that Ni-W is indeed capable of sustaining AIFs, with these features observed above ~1000 °C. The AIF formation also creates a new region of high temperature grain size stability that had not been previously observed.

20 μm thick nanocrystalline Ni-W films with 5-6 at.% W were created using the pulsed electrodeposition technique described by Detor and Schuh [49]. Deposition was performed on a

>98 wt.% pure Ni substrate which had been mechanically polished to a mirror finish using a diamond suspension prior to deposition. A plating solution consisting of sodium citrate, ammonium chloride, nickel sulfate, sodium tungstate and sodium bromide was heated to a temperature of 75 °C. Cycles of forward current density at 0.20 A/cm$^2$ for 20 ms, and reverse current density at 0.20 A/cm$^2$ for 3 ms were applied between the Ni substrate and a commercially pure Pt mesh, the insoluble counter-electrode, for 1 h. The samples were encapsulated under vacuum in high purity quartz tubes, suspended in a vertically-oriented tube furnace during annealing, and then plunged into a water bath in under 1 s for quenching. The samples underwent rapid quenching to preserve any thermodynamically-stable states that are only accessible at elevated temperatures.

X-ray diffraction (XRD) was performed using a Rigaku SmartLab X-ray Diffractometer operated at 40 kV and 44 mA with a Cu Kα source. Profiles were analyzed using the Rigaku PDXL2 software, with grain sizes determined using the Scherrer equation on the (111) peak. Transmission electron microscopy (TEM) samples were created using the focused ion beam (FIB) lift-out technique on an FEI Quanta 3D FEG Dual Beam and FEI Nova 600 scanning electron microscope (SEM)/FIB. To reduce ion beam damage, all TEM samples received a final polish with a low power 5 kV beam. The global film composition was verified in the SEM using energy dispersive X-ray spectroscopy (EDS) operating at 30 kV. Grain boundary compositional analysis was performed in a JEOL 2800 TEM at 200 kV and FEI Tecnai G(2) F30 S-Twin at 300k V using EDS in conjunction with scanning, bright field, and high resolution TEM to inspect the grain size and grain boundary structure. Phase identification was gathered under the same conditions using selected area electron diffraction (SAED). Grain size was calculated from bright field TEM images by measuring the areas of at least 50 grains and calculating the average equivalent circular diameter. Fresnel fringe imaging was used to identify interfacial films and ensure edge-on orientation of the grain boundaries during imaging [50].

First, a Ni-5 at.% W sample was annealed at 650 °C for 24 h, with the goal of coarsening the microstructure so that TEM samples only had one grain through the thickness, to allow for easy viewing of grain boundary structure. The sample was then heated to 1100 °C (~73% $T_{melting}$ for Ni-5 at.% W) for 10 min and then quenched. The grain structure is shown in bright field TEM images in Figures 1(a) and (b), where equiaxed grains are observed. No precipitates or second phases were detected in the film (Figure 1(c)). This is in agreement with the bulk Ni-W phase diagram, which predicts solid solution for compositions up to ~10 at.% W [51], even at high temperatures. The microstructure was further inspected using high resolution TEM and scanning TEM-EDS to analyze the grain boundary structure and composition. Multiple AIFs with an average thickness of approximately 1 nm are shown in Figure 2. In Figure 2(c), the image has been digitally enhanced using a fast Fourier transform bandpass filter to smooth background variations down to 2 nm, so as not to disturb the presentation of the grain boundary structure. It is important to note that the AIF thicknesses are all constant along the length of the interface, which is indicative of nanoscale intergranular films in equilibrium at the grain boundary. In contrast, wetting films tend to be much thicker and have an arbitrary thickness that is dependent on the amount of available liquid phase [41, 52]. Moreover, the annealing temperature is below the lowest eutectic on the Ni-W phase diagram, meaning it would be impossible for the grain boundary structure to come from a liquid phase. The EDS line scan across a grain boundary in Figure 3(b) with the associated scanning TEM image in Figure 3(a) shows increased W composition at the grain boundary compared to the crystalline interior. While absolute EDS compositional values are subject to spatial averaging due to the beam interaction volume, W segregation is clearly observed.

Only moderate W segregation is observed, which agrees with prior research where W was observed to be a weak segregator due to the significant solid solubility of W in Ni [36, 49]. The sample had an average grain size of 174 ± 54 nm, as shown in Figure 1(a) and (b). The high volume fraction of grain boundaries, approximately 1-2% [53], due to the grain size accommodates an elevated percentage of W dopant as reflected by Figure 3(b). While the grain interior composition is lower than 5 at.%, the significant volume of W-enriched grain boundaries accounts for the higher global average. If only the initial 650 °C annealing treatment is considered, our measured grain size would agree well with prior Ni-W annealing studies [36]. However, the primary differentiator here is the additional 1100 °C annealing treatment, which would be expected to cause rampant grain growth well beyond the measured value if trends from prior work were extrapolated. The absence of this rampant grain growth at very high temperature suggests that stabilization by AIFs may be occurring. To test this hypothesis, a detailed grain growth study was performed over a wide temperature range, going to higher temperature than had previously been studied.

A second set of similar Ni-W films were created with an initial average grain size of 35 ± 4 nm and a composition of 6 at.% W. Separate pieces of the film were then each subjected to 1 h annealing treatments at temperatures ranging from 200-1200 °C by inserting the sample at temperature followed by rapid quenching, with the resulting grain sizes plotted in Figure 4. Initially, negligible grain growth is observed up to ~600 °C, consistent with solute segregation serving as the primary stabilization factor to hinder grain growth [34, 54]. Annealing to 900 °C resulted in a 3-fold increase in the grain size, again consistent with previous observations [36, 55, 56]. However, the behavior that was observed at even higher temperatures was profoundly different than would be expected based on previous work. Further annealing to 1100 °C showed a secondary regime of stabilization with an average grain size of ~55-60 nm, indicating a significant deviation from expected grain growth trends. Finally, further annealing to 1200 °C yielded grain growth, with the grain size increasing rapidly.

A higher bulk composition that correlates to more segregated dopant can increase complexion width and prompt ordered-to-amorphous structural transitions [40], and also enhance thermal stability[36]. The Ni-6 at.% W alloy could have slightly thicker and a higher volume fraction of AIFs with improved nanocrystalline stability compared to 5 at.%, but these differences are expected to be small. Thermodynamic treatments of complexion formation have suggested that AIFs become the preferred grain boundary structure after a threshold transition temperature [40, 57]. For Ni-6 at.% W, we find that this transition temperature is between 900 and 1000 °C (~60-67% $T_{melting}$). Interestingly, the formation of AIFs in a Ni film doped with W is similar to segregation-induced grain boundary premelting that was previously observed in W-1 at.% Ni [58, 59]. Solid-state activated sintering observed in the W-rich system was caused by Ni segregation to form amorphous grain boundary complexions. In this study, even though W is a weakly segregating dopant in Ni, it appears that the boundary composition is enriched enough to enable premelting at some grain boundaries. AIFs have also been observed in Cu-Zr [47], Cu-Hf [44], Ni-Zr [44] and Mo-Ni [60] binary alloys, all of which have a negative $\Delta H^{mix}$ [61] and exhibit dopant segregation, corresponding to the general AIF material selection guidelines put forth by Schuler and Rupert [44]. In addition, AIFs have been observed in materials made by a wide variety of processing routes, having been observed in sputtered [44], ball milled [47], and sintered [59] materials. AIF formation is therefore a general phenomenon that can occur in a wide range of materials, including Ni-W where such features had not been observed previously. We suggest that AIFs have not been seen in prior works on Ni-W because either (1) heat treatments were not performed at high enough temperatures, (2) rapid quenching was not performed, or (3) detailed characterization of grain boundary structure was not performed.

In summary, electrodeposited Ni-W samples were analyzed to test the hypothesis that AIFs could form in this system. The Ni-5 at.% W sample demonstrated solute segregation and AIF formation, as well as a lack of dramatic grain growth even after annealing at 1100 °C. A similar sample set (electrodeposited Ni-6 at. % W), was annealed for 1 h at various temperatures up to 1200 °C, followed by rapid quenching, and two regions of grain size stabilization were observed. The first region coincides with solute segregation-enabled nanocrystalline stability which eventually transitions to grain growth, which has been reported on extensively in past work. Upon reaching temperatures sufficient for AIF formation, a second, unexpected region of nanocrystalline stability is observed. The observation that nanocrystalline Ni-W has smaller grain sizes at higher temperatures disrupts conventional thinking on thermal stability. Ultimately, this study indicates that it is possible to stabilize nanocrystalline grains at temperatures higher than previously expected through a new method that utilizes AIFs.

**Acknowledgements**

This research was supported by the U.S. Army Research Office under Grant W911NF-16-1-0369. XRD and TEM work was performed at the UC Irvine Materials Research Institute (IMRI). SEM and FIB work was performed at the UC Irvine Materials Research Institute (IMRI) using instrumentation funded in part by the National Science Foundation Center for Chemistry at the Space-Time Limit (CHE-0802913). Additional SEM and TEM work was performed at the Center for Integrated Nanotechnologies, an Office of Science User Facility operated for the U.S. Department of Energy (DOE) Office of Science by Los Alamos National Laboratory (Contract

DE-AC52-06NA25396) and Sandia National Laboratories (Contract DE-NA-0003525). The authors would like to thank Professor Mecartney of the University of California, Irvine and her research group for help with the 1200 °C annealing experiments.

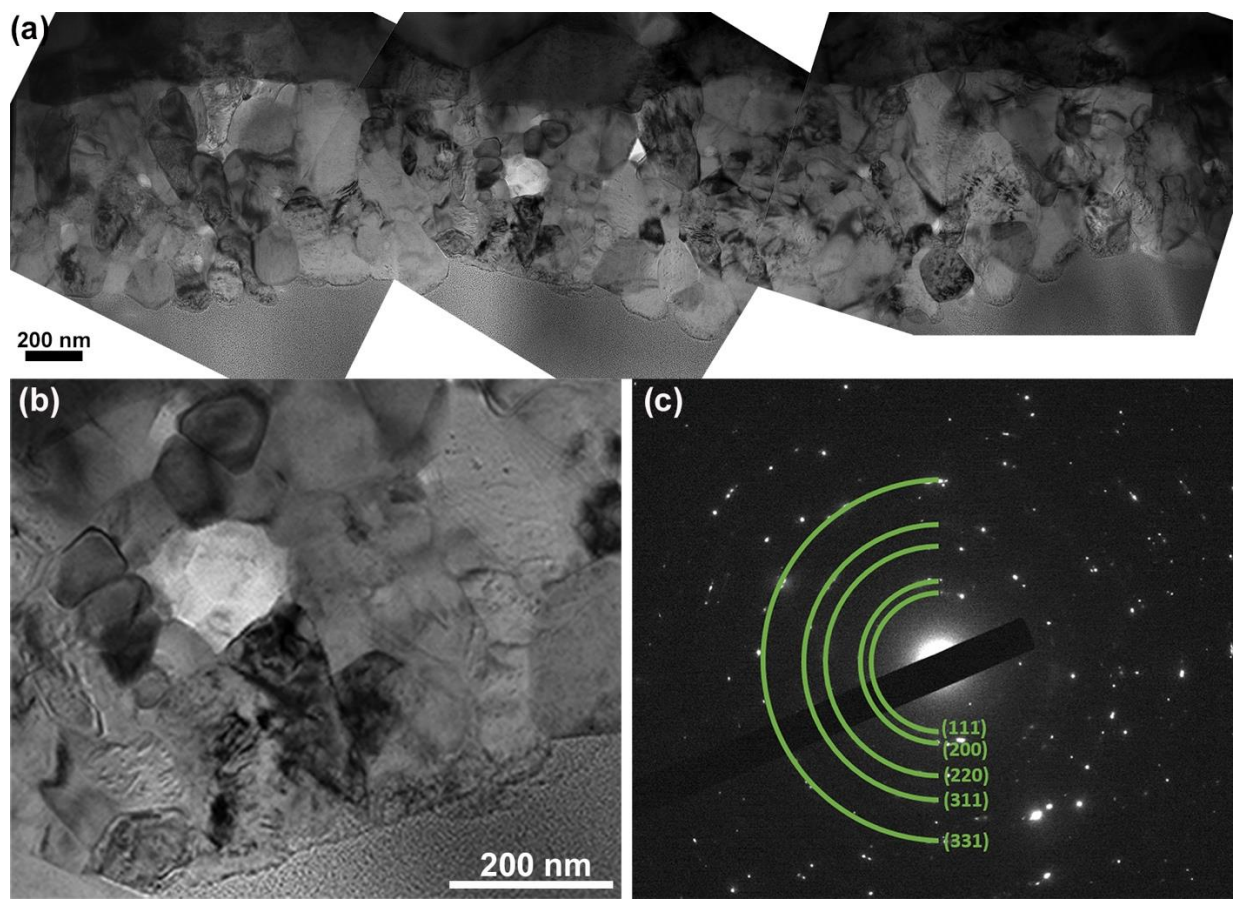

Figure 1. Bright field TEM images of nanocrystalline Ni-5 at.% W after a 24 h anneal at 650 °C and then a 10 min anneal at 1100 °C. (a) shows the overall film microstructure, while (b) shows a magnified image where equiaxed grains are observed. (c) The selected area electron diffraction pattern shows that only face centered cubic (fcc) Ni rings are present.

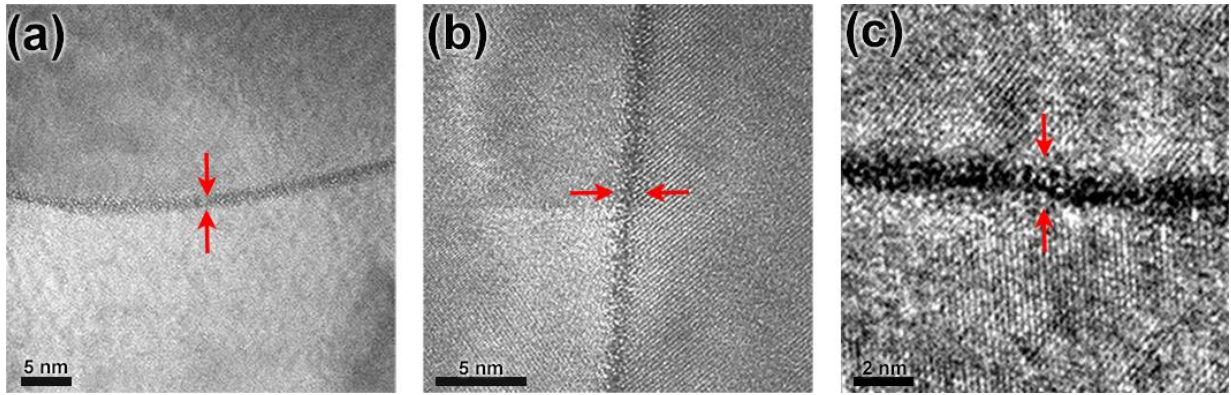

Figure 2. High resolution TEM images of ~ 1 nm thick amorphous intergranular films in the Ni-5 at.% W alloy after quenching from 1100 °C, with multiple examples shown in (a)-(c).

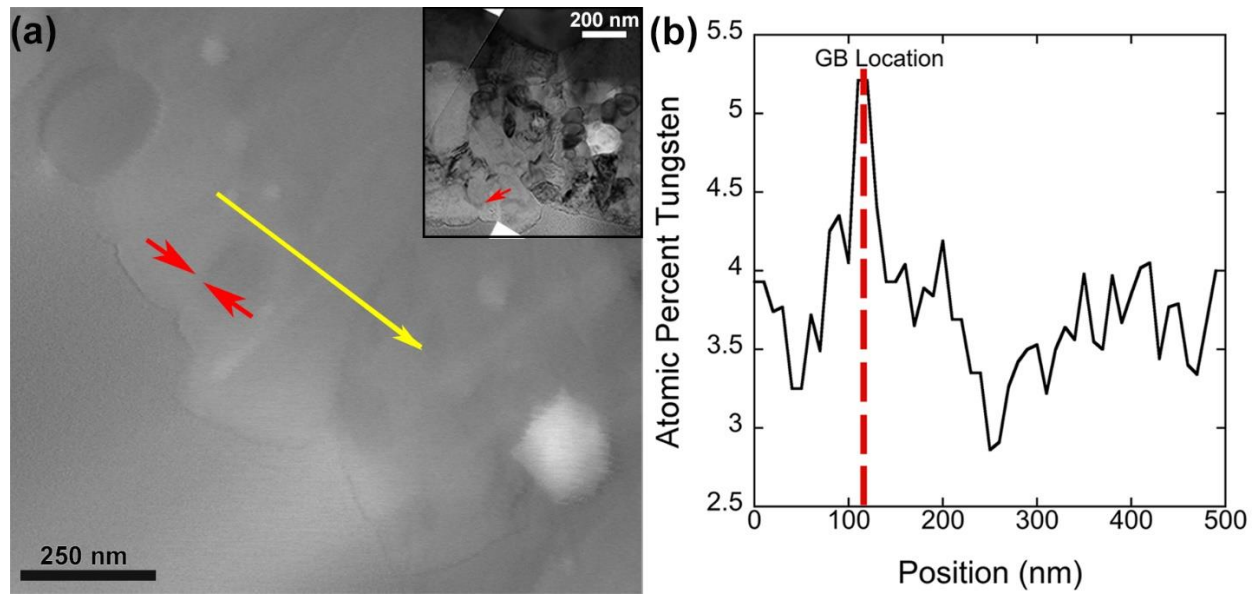

Figure 3. (a) Scanning TEM image of the Ni-5 at.% W alloy that was quenched from 1100 °C, with the associated bright field image shown in the inset. Red arrows in the scanning and bright field TEM images identify the boundary of interest, while the yellow line in (a) shows the scan location. (b) W concentration data from EDS demonstrates that W segregation occurs at the boundary.

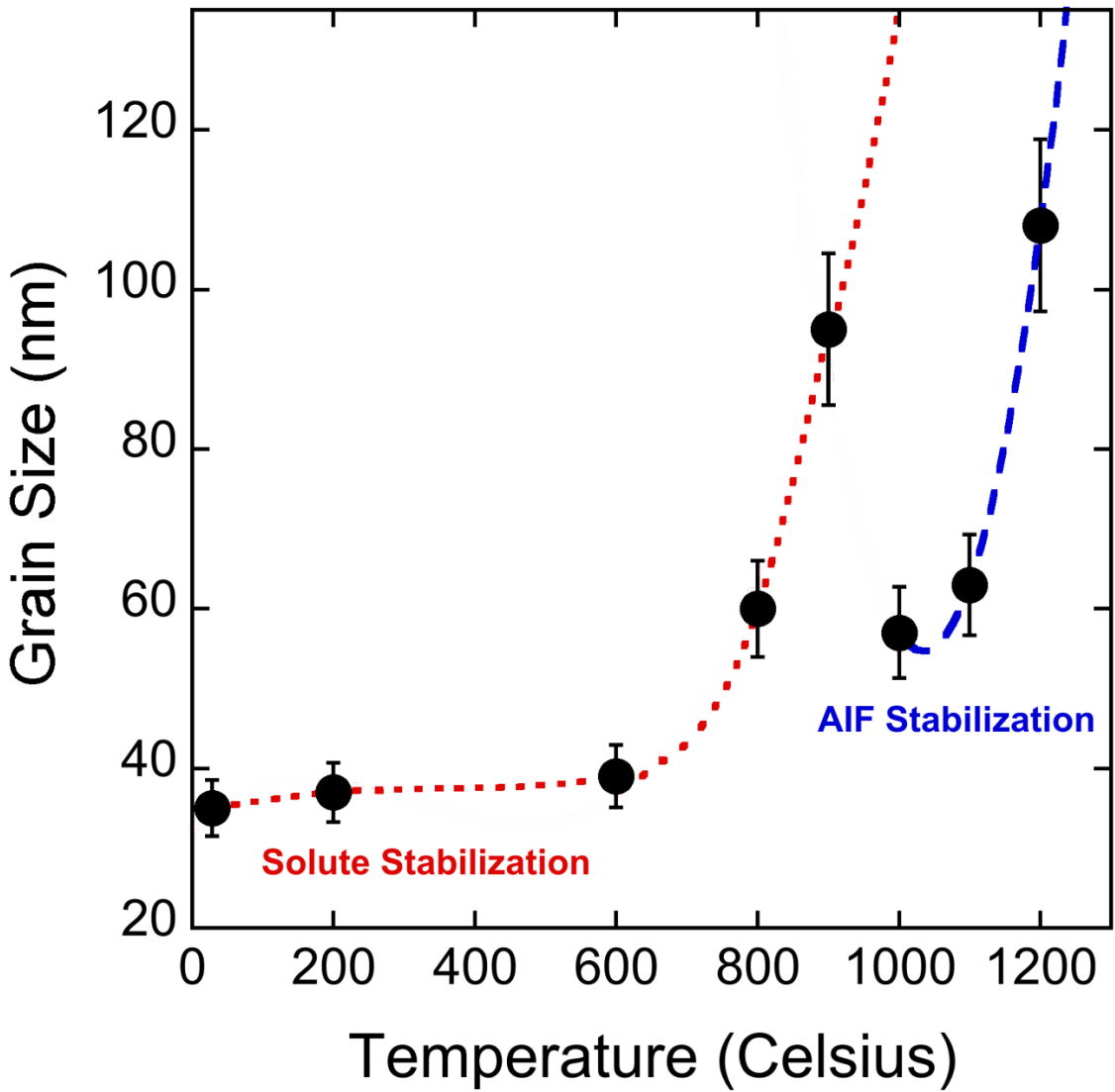

Figure 4. XRD grain size analysis of the Ni-6 at.% W electrodeposited alloys after annealing for 1 h at different temperatures. The data shows two distinct regions of stabilization by solute segregation and AIF formation. Rough trend lines are added to guide the eye and highlight these two regions.